\documentclass[12pt]{article}
\usepackage{color}
\usepackage[dvips]{graphicx}
\usepackage{dcolumn}
\usepackage{url}

\begin{document}
\Large
\begin{center}
  {\bf Computational Studies of Ruthenium and Iridium Complexes 
  for Energy Sciences and Progress on Greener Alternatives}
\end{center}
\normalsize

\vspace{0.5cm}

\noindent
Denis MAGERO\\
{\em School of Science, Department of Chemistry and Biochemistry,
Alupe University College, P.O.\ Box 845-50400, Busia, KENYA\\
e-mail: magerode@gmail.com}

\vspace{0.5cm}

\noindent
Tarek MESTIRI\\
{\em Laboratoire de recherche (LR 18ES19), Synth\`ese asym\'etrique et 
Ing\'enierie mol\'eculaire des Mat\'eriaux organiques pour l'Electronique 
organique, Facult\'e des Sciences de Monastir, Universit\'e de Monastir, 
5000 Monastir, TUNISIA\\
e-mail: mestiri.tarek@gmail.com}

\vspace{0.5cm}

\noindent 
Kamel ALIMI\\
{\em Laboratoire de recherche (LR 18ES19), Synth\`ese asym\'etrique et 
Ing\'enierie mol\'eculaire des Mat\'eriaux organiques pour l'Electronique 
organique, Facult\'e des Sciences de Monastir, Universit\'e de Monastir, 
5000 Monastir, TUNISIA\\
e-mail: kamealimi@yahoo.fr}

\vspace{0.5cm}

\noindent
Mark Earl CASIDA\\
{\em Laboratoire de Spectrom\'etrie, Interactions et Chimie th\'eorique 
(SITh),
D\'epartement de Chimie Mol\'eculaire (DCM, UMR CNRS/UGA 5250),
Institut de Chimie Mol\'eculaire de Grenoble (ICMG, FR2607), 
Universit\'e Grenoble Alpes (UGA)
301 rue de la Chimie, BP 53, F-38041 Grenoble Cedex 9, FRANCE\\
e-mail: mark.casida@univ-grenoble-alpes.fr} 

%

\vspace{0.5cm}
 
\noindent
This manuscript has been prepared for submission as a book chapter for
\begin{center}
Liliana MAMMINO (editor)\\
{\em Green Chemistry and Computational Chemistry},\\
a volume in the {\em Advances in Green and Sustainable Chemistry Series},\\
Due for publication July 2021,\\
Series list: {\tt https://www.elsevier.com/catalog/all/all/all/advances-in-green-chemistry}
\end{center}


\vspace{0.5cm}

\begin{center}
{\bf Abstract}
\end{center}
The energy sciences attempt to meet the increasing world-wide need for energy, 
as well as sustainability goals, by cleaner sources of energy, by new 
alternative sources of energy, and by more efficient uses of available energy. 
These goals are entirely consistent with the principles of green chemistry.
This chapter concerns devices for creating 
electricity from light and for creating light from electricity. The major 
focus is on the photoproperties of ruthenium and iridium complexes,
which have been proven to be a rich source of inspiration for conceiving 
photoactivated devices, including organic photovoltaic (OPV) cells and 
organic light emitting diodes (OLEDs). The chapter reviews important 
already-in-use and potential 
applications of ruthenium and iridium complex-based photodevices, including
the underlying mechanism behind their functioning and its investigation 
through computational chemistry approaches. It highlights the role of the 
information obtained from computational studies for the design of more 
efficient photodevices. The final part complements the discourse with a 
review of the progress on greener alternatives for OPVs and OLEDs.

\section{Introduction}
\label{sec:intro}

\begin{figure}
\begin{center}
\includegraphics[width=0.6\textwidth]{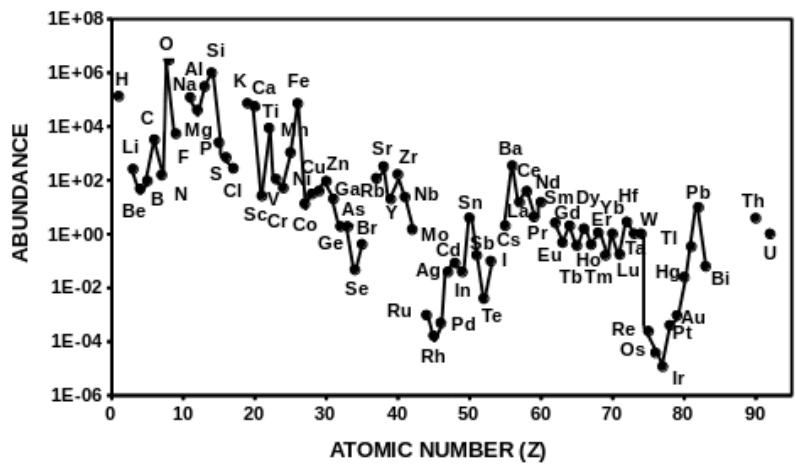}
\end{center}
\caption{
Relative abundance of different metals in the Earth's continental crust
in units of atoms of the element per 10$^6$ atoms of silicon.  Figure
adapted from Ref.~\cite{USGSF}.
\label{fig:abondance}
}
\end{figure}
Alternative energy sources and alternative
devices for making efficient use of energy fall nicely under the
objectives of green chemistry.  The most common source of energy nowadays 
remains the combustion of nonrenewable fossil fuels.  Undesirable
effects of fossil fuel combustion include air pollution contributing to 
global warming as well as political conflicts over the control of fossil
fuel deposits.  
One of the objectives of green chemistry is to reduce the consumption of
nonrenewable resources.  
Among the various alternatives to fossil fuels are
photovoltaics (PVs).  Solar energy is plentiful and solar cell (SC) 
technology promises to provide a ready source of relatively clean abundant
energy for decades to come.  Meanwhile light-emitting diode (LED) 
technology is revolutionizing how we light our homes and the displays
on our computers, television, and phones.  Although SCs and LEDs
may seem unrelated, they are roughly speaking the same device run either 
``forward'' to generate electrons from photons or run ``backward''
to generate photons from electrons.
This makes the two particularly natural to discuss in a single chapter.  
This chapter focuses on particular types of SCs and LEDs, namely dye-sensitized
solar cells (DSSCs) and organic LEDs (OLEDs).  This choice is governed both
by the authors' expertise and by the importance of these devices on the broader
field of PVs.  At the heart of conventional DSSCs and conventional OLEDs is 
another green chemistry problem, namely the use of such rare elements
(Fig.~\ref{fig:abondance}) as ruthenium (DSSCs) and iridium (OLEDs).
After discussing conventional DSSCs and OLED technology and how 
computational chemistry helps to understand atomistic aspects of underlying
processes and to create more efficient devices, still 
greener alternatives will be discussed that aim at eliminating rare
metals from DSSCs and OLEDs.

The rest of this chapter focuses on the atomistic understanding of luminescent
molecules as used in making photomolecular devices (PMDs \cite{SCC+94}) such as DSSCs
and OLEDs.  Such devices often rely on the heavy-metal effect of ruthenium
and iridium to enhance intersystem crossing probabilities.  Some of current
efforts to find greener solutions to replace rare metals will be reviewed.
Emphasis will be placed on how computational chemistry is helping PVs.
Section~\ref{sec:conclude} concludes.

\section{Photochemistry and Photodevices}
\label{sec:review}

This section provides a brief review of the science of photoprocesses
followed by engineering aspects which are important for reading the
scientific literature relevant to this chapter.  The principles in this 
section are meant to be general.  More system-specific principles will 
be discussed in subsequent sections.

\subsection{Photochemistry}
\label{sec:photochem}


Photochemistry is usually defined as the study of light-induced chemical
reactions.  However chemiluminescence (the emission of light by chemical
reactions) is also usually classed as photochemistry.  
Notice that there is
an obvious analogy between solar cells (electricity generated by light)
and photochemistry (in the sense of reactions generated by light) as 
well as an obvious analogy between light emitting diodes (light generated 
by electricity) and chemiluminescence, in part because of the electrochemical
nature of SCs and LEDs.  In fact, as we shall see, photovoltaics looks
much like photochemistry (in the most general sense) when examined at the
atomic scale.  Here we only touch on some basic concepts and vocabulary
from photochemistry.  No attempt is made to review all of photochemistry; 
we are only setting the stage for later sections.  For those seeking more
in-depth treatments of photochemistry, we may direct them to any one of a
number of excellent books \cite{MB90,T91,KM95,TRS10}.

\begin{figure}
\begin{center}
\includegraphics[width=\textwidth]{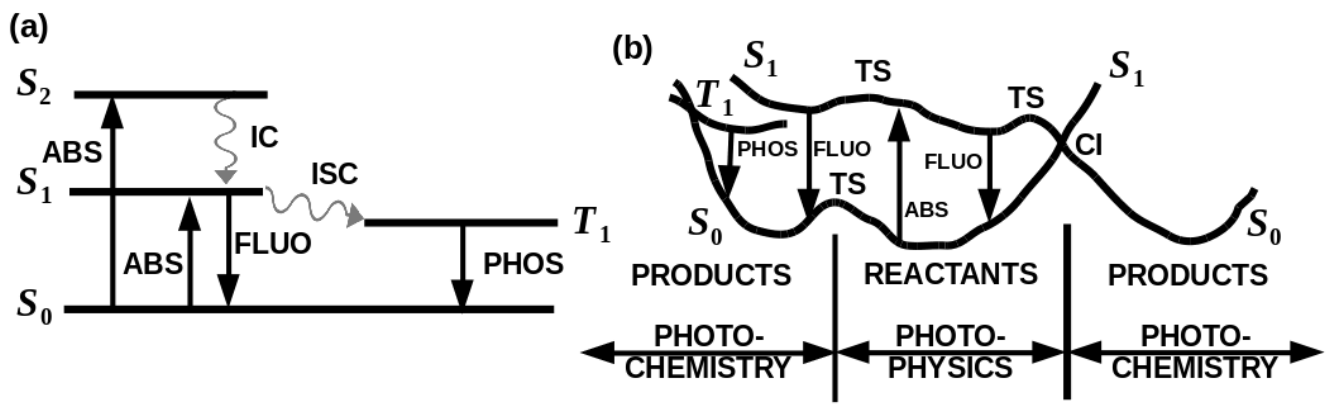}
\end{center}
\caption{
Two ways of representing photoprocesses: (a) Jablonski diagram,
(b) potential energy curves (1D PESs.)  Notation: $S_0$, ground 
singlet state; $S_1$, lowest excited singlet state; $T_1$, lowest 
triplet state; ABS, absorption; FLUO, fluorescence; PHOS, phosphorescence; 
ISC, intersystem crossing; CI, conical intersection; TS transition state.
\label{fig:PESa}
}
\end{figure}
As Fig.~\ref{fig:PESa} shows, photochemistry is a game played on several
levels.  Jablonski diagrams [such as Fig.~\ref{fig:PESa}(a)] represent
these levels by simple lines and photochemical processes by straight
arrows if light is absorbed or emitted and by wiggly lines if light is
emitted.  Modern photochemistry is based upon different potential energy 
surfaces (PESs), one for each electronic state (strictly speaking we
will be using the notion of adiabatic PESs though diabatic PESs are often
also used.)  
If only one degree of freedom
(e.g., some hypothetical reaction coordinate) is considered, then we have 
a 1D cross section (or potential energy curves) as shown in 
Fig.~\ref{fig:PESa}(b).  
Of the various processes shown in Fig.~\ref{fig:PESa}, the luminescent processes
are the most important for this chapter.  There are two of these, namely
(i) fluorescence (FLUO) which is a spin-allowed photon emission between states
of the same multiplicity from a short-lived excited state with a lifetime of
10$^{-12}$ to 10$^{-16}$ s and (ii) phosphorescence (PHOS) obtained after
a change of spin [intersystem crossing (ISC)] due to spin-orbit coupling
(SOC) to an excited state with longer lifetimes on the order of 10$^{-3}$ to 10$^{-2}$ s.

We will be concerned with the efficiency of different photoprocesses, such
as luminescence.  Let us consider a generic process $x$.  The quantum yield
$\Phi_x$ of $x$ is defined mathematically as,
\begin{equation}
   \Phi_{x} = \frac{\textrm{number of molecules undergoing process ($x$)}}
              {\textrm{number of photons absorbed}}
            = \frac{\textrm{rate of process ($x$)}}
              {\textrm{rate of light absorption}}
  \, .
   \label{eq:Denis.1}
\end{equation}
This is related to the rate constant $k_x$ of this process and to the
rate constants $k_y$ of this ($y=x$) and of competing ($y\neq x$)
processes, $\Phi_{x} = k_x/(\sum_y k_y)$.
Rates may alternatively be described in terms of lifetimes $\tau_x = 1/k_x$.
The lifetime of an excited state depends upon the rates of all the 
de-activation pathways, $\tau = 1/(\sum_y k_y)$.

\subsection{Solar Cells}
\label{sec:OPV}

\begin{figure}
\begin{center}
\includegraphics[width=0.5\textwidth]{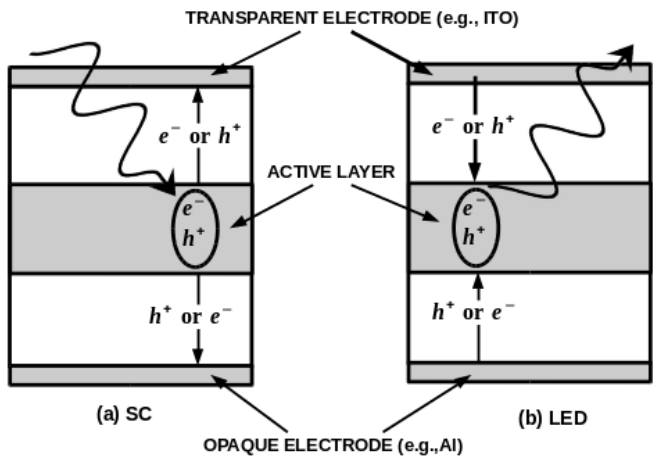}
\end{center}
\caption{
A schematic representation of how a solar cell (a) and how
a light-emitting diode (b) work.  The wiggly line represents light.
See text for a more detailed description.
\label{fig:SCandLED}
}
\end{figure}

Figure~\ref{fig:SCandLED}(a) shows a schematic representation of
an organic SC.  There are two electrodes, at least one of
which must be transparent to let light enter the SC.  A typical
transparent conducting material for this purpose is indium tin
oxide (ITO). These electrodes may be protected by additional layers
to prevent corrosion and/or the formation of unwanted electronic effects.
Sandwiched between the electrodes are a hole ($h^+$) conducting donor (D) layer 
and an electron ($e^-$) conducting acceptor (A) layer.  These may be solid
or stacked nanoparticles or even liquid, may be molecular or polymeric.  
Between the A and D layers is what we shall refer to as the active layer.  
This active layer may or may not be a different substance.  For example, 
in a polymer heterojunction SC, the active layer is simply a region a 
few nanometers thick which is close enough to the heterojunction for 
the exciton to arrive safely without undergoing $e^-/h^+$ recombination.  
This heterojunction may be simple in form or even take on a 
highly convoluted shape composed of spaghetti-like interpenetrating D and A 
layers.  In a Gr\"atzel (or Graetzel, as German allows ``\"a'' to be typed
as ``ae'') cell, the active layer 
consists of a dye adsorbed onto TiO$_2$ nanoparticles which form the 
A layer.  So, Fig.~\ref{fig:SCandLED} is indeed very schematic as it is
intended to cover a variety of cases.

The SC works by light creating an electron/hole ($e^-/h^+$) pair
(also known as an exciton) within the active layer.  The electron
then diffuses into the A layer and the hole into the D layer where
they make their way to their respective electrodes and generate
electricity.  In the case of particular types of SCs, each step of
this mechanism might be broken down into further steps, but this
simple description of the mechanism is adequate for present purposes.

\begin{figure}
\begin{center}
\includegraphics[width=0.8\textwidth]{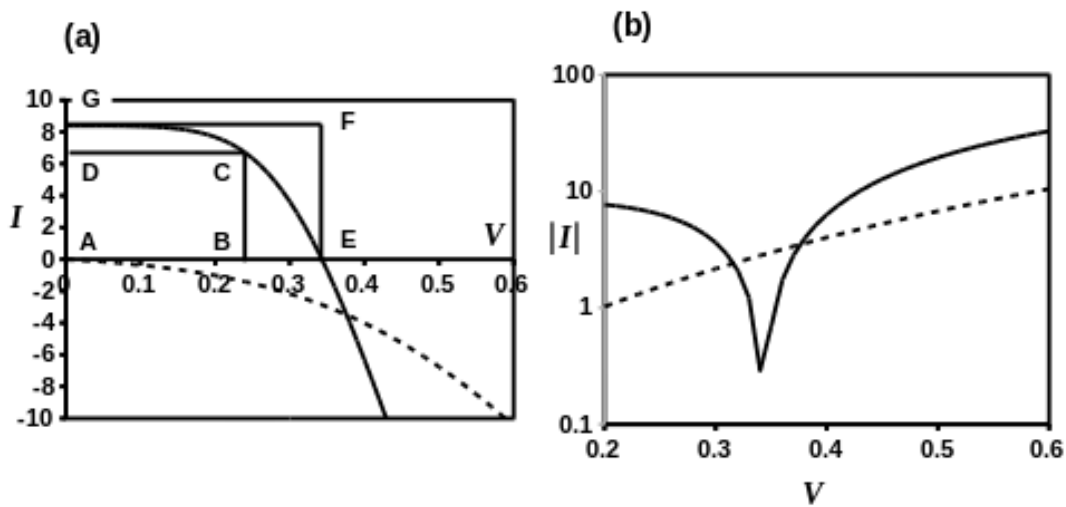}
\end{center}
\caption{
$I$-$V$ curves calculated with Eq.~(\ref{eq:intro.3}): solid curve 
(illuminated), dashed curve (without light.)  $V$ in volts; $I$ in $\mu$A.
The rectangles in part (a) are used to calculate the fill factor as described 
in the text.  Part (b) is an alternate representation of the same data
which is often used in the literature.  Parameters used in the calculation: 
$R = \mbox{0.01 M$\Omega$}$, 
$V_T = \mbox{0.1 V}$,
$n=1.5$,
$I_0 = \mbox{0.4 $\mu$A}$,
$I_{ph} = \mbox{8 $\mu$A}$.
\label{fig:IVcurve}
}
\end{figure}
Experimentally, the performance of a SC is characterized in terms
of the current, $I$, measured as a function of an applied voltage,
$V$, under illuminaton.  An $I$-$V$ curve  is obtained similar to
that shown in Fig.~\ref{fig:IVcurve}.
The fill factor, FF, is defined as the ratio of the area 
of the rectangle (ABCD) of maximal area that can be placed under
the $I$-$V$ curve to the area of the smallest 
rectangle (AEFG) containing the $I$-$V$ curve defined by the product of the 
short-circuit current, $I_{sc}$ (at $V=0$), with the open-circuit
voltage, $V_{oc}$ (at $I=0$), 
\begin{equation}
   \mbox{FF } = \frac{I_m V_m}{I_{sc} V_{oc}} \, , 
   \label{eq:intro.1}
\end{equation}
where $I_m$ and $V_m$ are the values at point C.
To quote Gr\"atzel \cite{G09}, ``The value of the fill factor reflects the
extent of electrical (Ohmic) and electrochemical (overvoltage) losses.''
If we also know the power density, $P_s$, of illumination, then we can
calculate the SC photoconversion efficiency (PCE), $\eta$, 
\begin{equation}
  \eta = \frac{I_{sc} V_{oc} \mbox{ FF}}{P_s} \, .
  \label{eq:intro.2}
\end{equation}
This formula can be shown to be equivalent to the ratio of the electrical
power generated by the SC to the light power absorbed.  It is usually
determined at AM 1.5G conditions, meaning using the solar spectrum at ground
level as seen through an air mass of 1.5 times that of the atmosphere
(to account for longer light paths when the sun is not directly overhead.)
Schokley and Queisser showed that $\eta$ has a thermodynamic upper
limit of 44\% \cite{SQ61}.  The modern accepted theoretical upper limit is
34\% at AM 1.5G.  Note that this value is only for simple SCs, as opposed
to, say, SCs created by stacks of simple SCs.  
Figure~\ref{fig:IVcurve} was constructed using the Schokley diode equation 
with an added photocurrent term $I_{ph}$,
\begin{equation}
  I = I_0 \left( 1 - e^{\frac{V+IR}{nV_T}} \right) + I_{ph} \, .
  \label{eq:intro.3}
\end{equation}
Here $V_T = kT/e$ is the thermal voltage, expressed in terms of Boltzman's
constant $k$, the temperature $T$ (in Kelvin), and the (absolute value of)
the charge of the electron $e$.  The diode ideality factor, $1 \leq n \leq 2$,
is introduced to take electron relaxation into account.  
The Schokley diode equation is a standard way to describe conventional
inorganic semiconductor devices \cite{SN07} but it is also widely used to 
describe organic devices \cite{JWK07}, although the underlying physics may be
very different than for inorganic devices \cite{GWWF10}.  Of course, 
LEDs and OLEDs are governed by similar equations.

The easiest quantity to model in Eq.~(\ref{eq:intro.2}) is the
open-circuit voltage, $V_{oc}$, at least under some idealized assumptions.  
In particular, Schokley and Queisser introduced
the two detailed balance assumptions that: (i) every photon whose energy 
is greater than the fundamental gap $E_g$ is absorbed, and (ii) every absorbed 
photon creates an electron-hole pair.  It can then be shown that,
\begin{equation}
  e V_{oc} \leq E_g =  \mbox{IP$_D$} - \mbox{EA$_A$} \, ,
  \label{eq:intro.9}
\end{equation}
where the fundamental gap $E_g$ is the difference between the electron affinity
of the acceptor, EA$_A$, and the ionization potential of the donor, IP$_D$.
Empirically it has been found in certain cases that,
\begin{equation}
  e V_{oc} = \mbox{IP$_D$} - \mbox{EA$_A$} - \mbox{constant} 
  \approx  (\epsilon^A_{\mbox{LUMO}} - \epsilon^D_{\mbox{HOMO}}) 
   - \mbox{constant} \, ,
  \label{eq:intro.11}
\end{equation}
where the constant reflects the nature of the electrode contacts and
band bending effects in the particular SC \cite{MBHR03}.  The orbital
energy approximation
in Eq.~(\ref{eq:intro.11}) is valid to the extent
that a Koopmans-like theorem holds \cite{DCT+15}.
Thus, within this approximation, $V_{oc}$ may be optimized by
designing the D and A components to have desirable relative
HOMO and LUMO energies.

\subsection{Light-Emitting Diodes}
\label{sec:LED}

Figure~\ref{fig:SCandLED}(b) shows a very schematic representation of
an LED.  It should be clear that an LED is ``just a SC run 
backwards,'' so that some of the same descriptive physics applies.
For example, plots such as that in Fig.~\ref{fig:IVcurve} are also
common in the LED literature but typically also include the radiance
of the LED on the same plot which is often roughly proportional to
the current.

The external efficiency of an LED may be described phenomenologically
by the equation,
\begin{equation}
  \eta_{\mbox{ext}} = \chi_{\mbox{out}} \beta \gamma \Phi_{\mbox{PL}} 
  \, , 
  \label{eq:LED.1}
\end{equation}
where $\chi_{\mbox{out}}$ is the optical coupling factor which describes
light loss due to the interface with the transparent electrode,
$\beta$ decribes the fraction of excitations that emit light,
$\gamma$ is the charge balance factor which describes the ratio of 
charge recombination to charge injection, and $\Phi_{\mbox{PL}}$
is the photoluminescence (PL) quantum yield \cite{YRC+11}.  If the interest
is in, say, interior lighting, then multiplication by the color
rendering index (CRI) may also be appropriate, though it is not included
in Eq.~(\ref{eq:LED.1}).  The two factors of greatest importance for us
are $\beta$ and $\Phi_{\mbox{PL}}$, though we will also say a word
about $\gamma$ at the appropriate point below.  The PL 
efficiency is related to radiative versus nonradiative factors.
In particular, radiative rates ($k^r$) should be maximized
relative to nonradiative rates ($k^{nr}$) so as to maximize,
$\Phi_{\mbox{PL}} = k^r/(k^r+k^{nr})$.

\begin{figure}
\begin{center}
\includegraphics[width=0.6\textwidth]{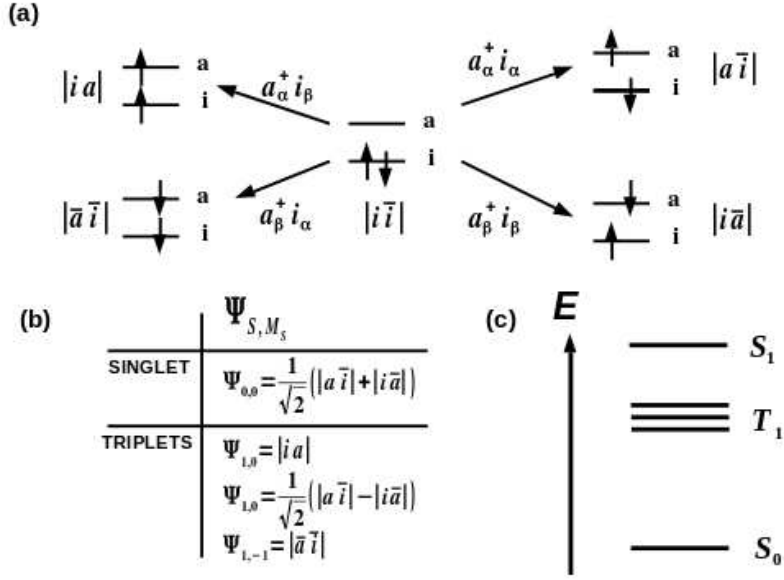}
\end{center}
\caption{
Two-hole two-electron model (TOTEM): (a) ground and singly excited
states, (b) spin-adapted wave functions, and (c) Jablonski 
diagram.  See text for additional explanation.
\label{fig:TOTEM2}
}
\end{figure}
Let us consider what happens when the electron and the hole meet in the
active layer.  To do this, we will consider the two-hole two-electron 
model (TOTEM) shown in Fig.~\ref{fig:TOTEM2}.  There are two orbitals.
The lower-energy orbital $\psi_i$ is doubly occupied and the higher-energy 
orbital $\psi_a$ is empty in the ground state (GS).  Simultaneous
hole migration into $\psi_i$ and electron migration into $\psi_a$
leads to the four single determinant wavefunctions shown in part (a)
of the figure.  These are not all true excited states (ESs) as not all
of them are spin eigenfunctions.  Solving the Schr\"odinger equation 
in the absence of spin-orbit coupling leads to the spin-adapted wave
functions shown in part (b) of the figure.  Relative to the $S_0$
GS, the corresponding open-shell singlet ($S_1$) and triplet ($T_1$) 
ES energies are, $\Delta E_S  =  \epsilon_a - \epsilon_i + 2 [ai \vert ia] - [aa \vert ii]$, 
where $\epsilon_i$ and $\epsilon_a$ are orbital energies, 
$[aa \vert ii]$ is a Coulomb repulsion integral, and 
$[ai \vert ia]$ is an exchange integral.  
We have the classic result that the corresponding open-shell singlet
ES is higher in energy than the triplet.  This leads to the Jablonski
diagram shown in part (c) of the figure.  As photons carry no spin,
radiation selection rules predict that fluoresence 
($S_1 \rightarrow S_0$) is allowed, while phosphorescence 
($T_1 \rightarrow S_0$) is forbidden.  We can now understand
the most fundamental problem in LED theory which is that 
the fraction of excitations that emit light nominally is only 
$\beta = 1/4$ in Eq.~(\ref{eq:LED.1}) in the absence of spin-orbit
coupling.  We shall call this the spin-statistics problem. 
It also leads to a lowering of the PL efficiency. 
Altogether the result is an unacceptable inefficiency in LEDs.  
Later, in Secs.~\ref{sec:IrOLED} and \ref{sec:MF-OLED}, we will discuss
various strategies used to overcome the spin-statistics problem. 

\section{Heavy Metal Complex Photoprocesses}
\label{sec:heavymetal}

At the heart of every photodevice are atomistic-level photoprocesses.  For the 
cases discussed in this section and in the next section (Sec.~\ref{sec:green}),
namely dye-sensitized solar cells (DSSCs) and organic LEDs (OLEDs), there are
specific molecules that capture light to produce electrons or capture electrons
to produce light.  The former are examples of photochemical molecular 
devices (PMDs) which have been defined as structurally organized and 
functionally integrated systems capable of elaborating the energy and 
information input of photons to perform complex functions \cite{SCC+94}. 
PMDs, DSSC dyes, and sensitizers for OLEDs may all be viewed as particular cases 
of molecular electronics, whose origin is usually attributed to Aviram and Ratner
in their famous 1974 article  \cite{AR74}.  Because of lack of practical applications,
molecular-level electronics is sometimes viewed as only an emerging field,
and yet we are about to delve into some very practical present-day examples of 
molecular electronics!

\subsection{Traditional Model}
\label{sec:tradmod}

The most frequently used dye in DSSCs is some form of (pseudo-)octahedrally-coordinated
polypyridinal ruthenium(II) complex and a (pseudo-)octahedrally-coordinated 
polypyridinal iridium(III) complex is commonly introduced into the active layer of 
organic LEDs (OLEDs).  The traditional model for describing the electronic structure
of the ground and excited states of such complexes is, of course, ligand field
theory (LFT) \cite{FH00}.  Even though {\em ab initio} and density-functional methods
have evolved to the point that they may now be applied to these complexes, LFT cannot
be avoided because it remains the language used in discussing these complexes.
In fact, one problem is how to recover an LFT picture from DFT so that computational
chemists using DFT can speak with experiments in a familiar LFT language.

The $O_h$ LFT splitting of the 5 $d$ orbitals into three lower $t_{2g}$ and two
higher $e_g$ orbitals is familiar from most university first-year chemistry courses.
As we will typically be dealing with high-field splitting, the expected 
GS LFT configuration is $t_{2g}^2 e_g^2$.  The $t_{2g}$ orbital is nonbonding with
polypyridinal ligands, but the $e_g$ orbital is actually antibonding.  Different from 
in the first-year courses, we will emphasize this by writing $e_g^*$.
Polypyridinal ligand $\pi^*$ orbitals are energetically interspersed between the
$t_{2g}$ and $e_g^*$ orbitals.

\begin{figure}
\begin{center}
\includegraphics[width=0.8\textwidth]{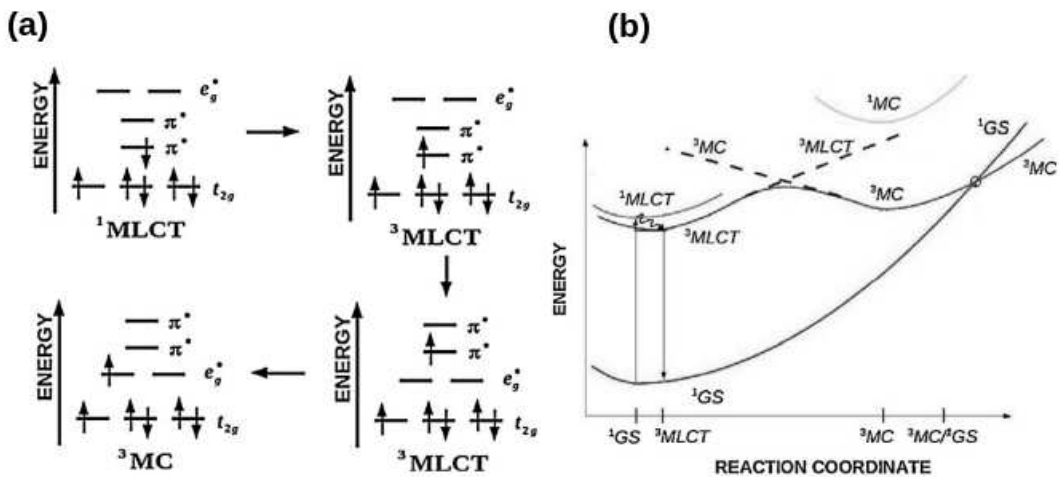}
\end{center}
\caption{
(a) Ligand field theory diagrams for the electronic ground state and the most
relevant excited state of a pseudo-octahedral $d^6$ complex.
Adapted from Ref.~\cite{MCA+17}.
(b)
The principle PES expected from LFT.  The dashed lines indicate
diabatic states whose avoided crossing leads to the energetic barrier
on the adiabatic surface between the $^3$MLCT and $^3$MC minima.
Adapted from Ref.~\cite{MCA+17}.
\label{fig:LFT}
}
\end{figure}
We are now in a good position to be able to understand the ESs 
and associated photoprocesses of our complexes [Fig.~\ref{fig:LFT}(a)]
\cite{SCC+94,WF11,MMA+16}.  They are, of course, governed by the same general
principles already described in Sec.~\ref{sec:photochem}. 
When a photon is absorbed by the singlet GS ($^1$GS),
the spin-selection rule tells us that the initial excited-state will also be
a singlet.  Furthermore the Laporte selection rule forbids $d \rightarrow d$
metal centered (MC) transitions, so we should expect the initial excited state
to be of metal-ligand charge-transfer (MLCT) character.  This may not be the
same $^1$MLCT ES shown in the upper right-hand side (RHS) of Fig.~\ref{fig:LFT}(a),
but radiationless relaxation will soon bring us down to this latter 
$^1$MLCT($t_{2g},\pi^*$) state.  No great change in geometry is expected
at this point.  However SOC due to the heavy-metal effect \cite{W12,XH14}
means that ISC can occur to populate the corresponding triplet state 
$^3$MLCT($t_{2g},\pi^*$)
[lower RHS of Fig.~\ref{fig:LFT}(a)].  If this is close enough to the $^3$MC($t_{2g},e_g^*$)
state, then configuration mixing will occur, allowing a transformation of the
$^3$MLCT character of the ES to $^3$MC character [lower left-hand side (LHS)
of Fig.~\ref{fig:LFT}(a)]. As the $\pi^*$ orbital is antibonding only within the 
ligand but the $e_g^*$ orbital is antibonding between the ligand to the metal, 
the $^3$MLCT state is expected to have approximately the same geometry as the $^1$GS 
but the $^3$MC state is expected is expected to have a longer metal ligand bond
length in comparison with the $^1$GS.  All of this is illustrated in Fig.~\ref{fig:LFT}(b)
where we also see the possibility of a triplet transition state ($^3$TS) between
the $^3$MLCT and the $^3$MC state.  Also shown in Fig.~\ref{fig:LFT}(b) is the
crossing of the $^1$GS and $^3$MC curves at large metal-ligand bond distance.
At this point, partial dissociation of the complex and/or interaction with the
solvent, combined with the heavy-metal effect, allows radiationless relaxation of
the $^3$MC state back to the $^1$GS.  Not shown in Fig.~\ref{fig:LFT}(b) is the
expected Jahn-Teller distortion of the MLCT and MC states due to the electronic
degeneracy (or near degeneracy) of the orbital configuration.
This LFT picture tells us that our complexes will have long-lived 
(phosphorescent) excited
states because of the spin-forbidden nature of the $^3$MLCT $\rightarrow$ $^1$GS
transition.  
\begin{figure}
\begin{center}
\includegraphics[width=0.4\textwidth]{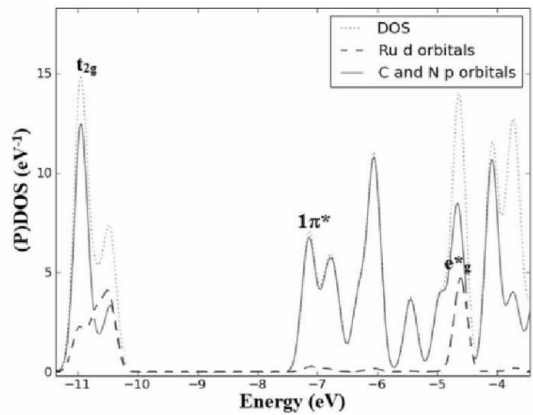}
\end{center}
\caption{
PDOS for a polypyridinal ruthenium(II) complex.
Adapted from Ref.~\cite{MCA+17}. 
\label{fig:PDOS}
}
\end{figure}
We still have the problem of recovering a LFT picture from
more sophisticated calculations.  Quantitatively correct geometries may be obtained
from DFT and TD-DFT is commonly used to calculate absorption spectra, but a connection
to LFT is still needed.  This is not really a problem for the nonbonding $t_{2g}$
orbitals which can often be readily identified by visualizing the molecular orbitals 
of the complex but the same method rarely allows identification of the antibonding
$e_g^*$ orbitals because they are heavily mixed with ligand orbitals.
Nevertheless we have found that a LFT-like picture may
be recovered from DFT calculations using the partial density of states (PDOS)
method \cite{WJL+14}.  Figure~\ref{fig:PDOS} shows (dotted line) the total density of 
states, 
\begin{equation}
  \mbox{DOS}(\epsilon) = \sum_i g(\epsilon-\epsilon_i) \, ,
  \label{eq:heavymetal.1}
\end{equation}
of a polypyridinal ruthenium (II) complex. Here the $\epsilon_i$ are the energies
of the occupied and unoccupied orbitals and $g$ is a normalized gaussian broadening 
function ($\int g(x) \, dx = 1$).  The partial DOS (PDOS) uses Mulliken population
analysis to assign how much of the DOS is due to metal $d$ orbitals and how much is
due to ligand $\pi$ and $\pi^*$ orbitals.  (The concepts of DOS and PDOS are
extensively discussed in the context of surface chemistry in Ref.~\cite{H88}.)
Specific formula for the PDOS are given 
in the supplementary material associated with Ref.~\cite{WJL+14}.  Suffice it to 
say that the gaussian broadening combined with the Mulliken population analysis
allows us a clear picture (shown in Fig.~\ref{fig:PDOS}) of where the 
$t_{2g}$ and $e_g^*$ orbitals lie, even though some inevitable fuzziness remains
due to the necessary choice of the full width at half maximum (FWHM) of the gaussian
broadening and some basis-set-dependence in the Mulliken population analysis.
It should also be noted that the $e^*_g$-$t_{2g}$ energy difference calculated
in this LFT-like approach is not the same as that in true LFT but still provides
a useful LFT-like picture.
We will return to the use of the PDOS in Sec.~\ref{sec:luminescence} when we
discuss an orbital-based luminescence index. 

\subsection{Ruthenium Complexes in Dye-Sensitized Solar Cells} 
\label{sec:PMD}


The first DSSC was that of O'Regan and Gr\"atzel
in 1991 \cite{OG91}.
In 2017, Michael Gr\"atzel received the Global Energy Prize ``for transcendent
merits in development of low cost and efficient solar cells, known as `Graetzel
cells' '' \cite{Graetzel17} or DSSCs.  
DSSCs are composed of a sensitizing dye that is adsorbed at the surface 
of nanoparticles made of a wide band-gap semiconductor [titanium oxide (TiO$_{2}$) in 
the original DSSC]. The dye absorbs light and transfers an electron to the 
semiconductor conduction band. The hole left in the dye is regenerated by a solution 
redox electrolyte [a triiodide/iodide (I$_3^-$/I$^-$) redox couple in the orginal
DSSC] or a solid hole conductor  and finally the electron migrates through the external 
load to complete the circuit \cite{G03,G09,FSR11,RP19}.   
The nanoparticulate nature of the semiconductor is critical as it provides a large surface
area for grafting a large amount of dye \cite{OG91,S94,SBA94,KG96,DWS+04,HTTL06,KDW07,XML+08}.
Charge separation occurs (in $n$-type DSSCs) through photo-induced electron injection 
from the dye into the conduction band of the solid nanoparticles \cite{G03,G09,FSR11,RP19}.  
Evidently the $S_1$ excited state of the dye must be higher in energy than the oxide
conduction band (CD) and the electrolyte redox couple must also be able to replentish
the dye with a new electron \cite{OH09,P10}.

The original 1991 DSSC had a 7.9\% efficiency under a light intensity of 8.3 mW.cm$^{-2}$ 
(AM 1.5G) and of 7.12\% under a light intensity of 75 mW.cm$^{-2}$ (AM 1.5G).
Much work has been done to improve this efficiency by working on the different DSSC
components \cite{WLC+12,HBS+10,NT09,MFB09,PJP10}.
The dye in the DSSC assembly plays a particularly significant role in the harvesting of 
solar energy and its conversion into electrical energy. 
Consequently much work has been done on improving the dye 
by various strategies ranging from incorporating functionalized ancillary
ligands such as triazolylpyridine to replacing the thiocyanate ligands with other chelating
anions \cite{YWW+15}.  The result has been has been consistent
improvement in efficiency to more than 11\% using {\em tris}(2,2-bipyridyl)ruthenium(II)
analogues \cite{GHD+18}.  One of the very best dyes has turned out to be 
[Ru(tcterpy)(NCS)$_3$] (tcterpy=4,4',4"-tricarboxy-2,2':6',2"-terpyridinne),
also known as ``N749'' or as the ``black dye'' because of its ability to absorb a remarkably
large part of the solar spectrum \cite{NPR+01,WYSA05,SSO+12,TSO+14}.

At the present time, DSSCs have been commercialized and are being sold by a number of
companies, including Dysol, Solaronix, SolarPrint, G24innovations, Sony Corporation,
H.Glass, and Exeger Operations AB.  The energyglass fascade of the Science Tower 
in Austria is the largest installation so far of DSSCs and was manufactured by 
H.Glass.



\subsection{Iridium Complexes in Organic Light Emitting Diodes}
\label{sec:IrOLED}


The basic theory of LEDs was already presented in Sec.~\ref{sec:LED}.
Thanks to work recognized by the 2000 Nobel Prize in Chemistry which was 
awarded to Alan J.\ Heeger, Alan G.\ MacDiarmid,
and Hideki Shirakawa ``for the discovery and development of conductive
polymers'' organic LEDs (OLEDs) are now possible.
These provide efficient light sources for lighting and displays.
Expertise in organic chemistry
means that OLEDs are arguably easier to fine tune and can be made more
cost efficiently.  In particular, this helps in engineering color-tunability, 
colour quality,  luminance efficiency, large-area display, wide-viewing angle, 
and lightness.  Also, OLEDs have the advantage over conventional LEDs in
that they can be made flexible \cite{A14}.
For all of these reasons, OLEDs are now a well-established commercial success
story.  This section describes conventional OLEDs, based upon the ``heavy atom
effect.''
(Reviews may ve found in Refs.~\cite{RHD+07,W07,YRC+11}.)


These OLEDs make use of the so-called ``heavy atom 
effect''---that is, the presence of
significant spin-orbit coupling (SOC) due to the presence of heavy atoms.  
This leads not only to lifting of the
degeneracy between the triplet states but also to singlet-triplet
mixing and
allows rapid conversion of the singlet into a triplet through
ISC.  At the same time, a singlet component
is introduced into the triplet states so that they may radiate by
phosphorescence. The resultant OLED is thus a phosphorescent OLED.
Care must be taken in ligand design to minimize nonradiative
decay.  One way to do this is by assuring short phosphorescence
lifetimes.  This is also important in designing OLEDs with values
of $\gamma$ [Eq.~(\ref{eq:LED.1})] close to unity when operating
at high power so that energy is dissipated quickly before other
processes can take place.  The rest of this section is devoted to
describing the particular role of the different electronic states
involved when metal complexes are used in OLEDs.  The reader interested 
in further information about this classic type of OLED is referred to 
Refs.~\cite{RHD+07,W07,YRC+11}.


As the SOC constant ($\zeta$) increases as the fourth power 
of the effective nuclear charge Z, 
then the 
most effective elements for introducing SOC are the heavy elements. 
Of these, the heavy metal ion iridium(III) is generally considered the most important 
OLED emitter \cite{RHD+07}. 
Indeed, the conventional solution to the spin-statistics problem dates back to 1999
when iridium phenylpyridine complexes were introduced into the active 
layer \cite{BLB+99}.  These OLEDs make use of the so-called ``heavy atom 
The high $\zeta$ values for iridium(III) 
and the large choice of ligands whose energy levels mix well with 
atomic orbitals of iridium(III) has necessitated the development of materials 
that emit with very high efficiency from triplet states.  
Not only do iridium(III) complexes emit from states with a large degree of 
MLCT character \cite{LSB+06}, but iridium(III) complexes 
are also highly emissive at room temperature, possess radiative lifetimes 
in the range of microseconds, which are advantageously shorter than the 
typical lifetimes of pure organic phosphorescent materials \cite{GPB16}. 
This in turn improves the quantum efficiency of OLEDs to values as high 
as 100{\%}.

\begin{figure}
\begin{center}
\includegraphics[width=0.6\textwidth]{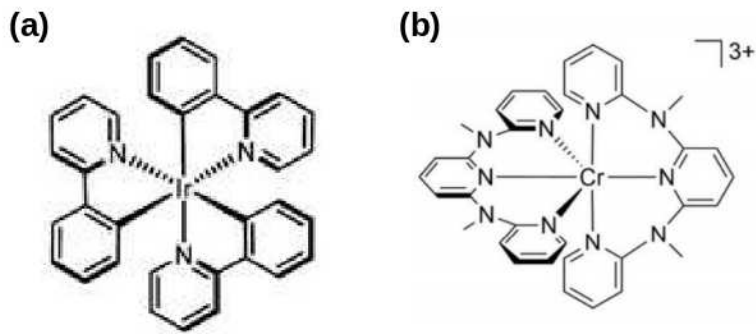}
\end{center}
\caption{(a) {\em Fac}-tris-phenylpyridineiridium(III), {\em fac}-Ir(ppy)$_3$, ppy = 2-phenylpyridine.
(b) ``Molecular ruby'':[Cr(ddpd)$_2$]$^{3+}$ \cite{OGF+15,ONE+17,OSB+17,OHHR18}.
\label{fig:LEDs}
}
\end{figure}

Iridium  complexes have already been used successfully in the design of 
efficient phosphorescent emitters in OLEDs 
\cite{SZ19,CCG+17,YJJ+17,ABF00,LSB+06,LHS+09,SDT+09,YRC+11,SRB+12,LCB+16,PSN+16,ZBP16,ZBP17,ZP18}. 
In particular, cyclometallated iridium(III) complexes have 
and are still in use as an OLED phosphors.  The archetypal structural 
unit of an iridium(III) complex comprises an aryl heterocycle, often a 
2-phenylpyridine unit, that is bound to the metal through the heteroatom 
of the heterocycle and through a metallated carbon atom  {\em ortho} to the 
interannular bond \cite{FBS+07}. An example is {\em fac}-Ir(ppy)$_{3}$ 
shown in Fig.~\ref{fig:LEDs}(a). 

%
%
%
%
%
%
%

\subsection{Modeling Luminescence}
\label{sec:luminescence}


Luminescence plays a key role in the theory of both DSSCs and OLEDs.  In the
DSSC case, the longer the luminescence lifetime, the greater the chance of transferring
an electron to the nanoparticulate semiconductor.  In the OLED case, we wish to
obtain as much light as possible through phosphorescence.  
The traditional model for luminescence in 
octahedrally-coordinated polypyridinal ruthenium(II) and iridium(II) complexes
has been explained in Sec.~\ref{sec:tradmod}.  It is used both as an explanation
of luminescence and as a hypothetical model for designing complexes with long-lived
luminescent states even though the basic model 
was intially developed for
[Ru(bpy)$_3$]$^{2+}$.  Our goal in this section is to go beyond [Ru(bpy)$_3$]$^{2+}$
and to validate this model, if possible, for the entire class of octahedrally-coordinated 
polypyridinal ruthenium(II) complexes.  As will soon become clear, this is an ambitious 
goal which requires us to make some drastic compromises.  Nevertheless the result will 
be an orbital-based luminesence index (LI) which seems to work remarkably well \cite{MCA+17}.

Experimental luminescence lifetimes measure the rate of disappearance
of the excited state {\em either} by luminescence {\em or} by other 
mechanisms such as radiationless relaxation.  Hence competing radiationless
relaxation routes can lead to significant reductions in luminescence
lifetimes.  According to the traditional model (Sec.~\ref{sec:tradmod}), 
we are particularly interested in $^3$MLCT $\rightarrow$ $^3$MC
pathway shown in Fig.~\ref{fig:LFT}(b) and whose rate is determined by the height
of the $^3$TS barrier.  Unfortunately finding the value of the height of this barrier is not
easy to do experimentally.  It involves measuring luminescence lifetimes at
many temperatures and then fitting them to an elaborate emiprial form  \cite{JBB+88}.
Moreover sufficiently complete temperature information for empirical fitting is only
available for a few compounds, one of which is tris(bipyridine)ruthenium(II) 
[Ru(bpy)$_3$]$^{2+}$ in propionitrile/butylnitrile (4:5 v/v) where the height of 
the $^3$TS barrier is 3~960 cm$^{-1}$ \cite{JBB+88}.

Another way to determine the height of the $^3$TS is theoretically using DFT.
This is not as straightforward as it may at first sound.  In general, the triplet
surface exhibits multiple different $^3$TSs.  This is especially true in the case
of a complex with several different types of ligands, but is also true in the case
of a complex with a single type of ligand.  In the later case, there are not only 
barriers for removing ligands but also barriers for pseudorotations between different
Jahn-Teller distorted geometries.  To find the right $^3$TS, it is first necessary to 
have a guess as to how the reaction might proceed.  Here we may be guided by
Adamson's rules which say that (1) the weakest bond will break if one bond is weaker
than the others or else (2) two bonds will break {\em trans} to each
other in the case that all bonds are of similar strength \cite{VC83}.
What is typically done once a pathway is decided upon is to vary some
representative coordinate while relaxing all other coordinates (``do a scan'')
to find a possible TS.  Convergence to this TS may be tightened using special
methods and vibrational frequencies are calculated to insure that it is indeed
a real TS with a single internal vibrational frequency.  In the past, some authors
made the error of identifying the scan path with the reaction path.  However
this is quite incorrect \cite{M80}.  Instead the intrinsic reaction coordinate (IRC) 
is calculated by following the PES down in energy on each side towards the reactants 
and the products.  Only then can the true activation energy of the reaction be
determined.  Besides the much studied [Ru(bpy)$_2$]$^{2+}$
complex \cite{DHAE17,SDV+17,SDAH18}, Sun {\em et al}.\ have determined the $^3$TS and activation
energy for two ruthenium(II) compounds ([Ru(mpy)$_2$]$^{2+}$ and [Ru(mphen)$_2$]$^{2+}$)
\cite{SDV+17}.  The results require careful examination as they are subject to variations
upon choice of functional and variations in the quality of the basis sets.  The
gas phase theoretical values of the $^3$TS activation energy for {\em cis} 
dissociation (1~364 cm$^{-1}$ \cite{DHAE17} and 1~136 cm$^{-1}$ \cite{SDV+17}) and for
{\em trans} dissociation (700 cm$^{-1}$ \cite{SDAH18}) of [Ru(bpy)$_3$]$^{2+}$ may be compared
with the above mentioned experimental value of 3~960 cm$^{-1}$ in solution.
It should also be kept in mind that ``chemical accuracy'' is at best
1 kcal/mol = 350 cm$^{-1}$ and that ``transition metal accuracy'' \cite{FMC+04}
is more like 5 kcal/mol = 1~700 cm$^{-1}$.

As trends are more important than absolute numbers, we decided to take
another approach to extracting a $^3$TS energy from experiment.  The
review article of Juris {\em et al.} \cite{JBB+88} provides almost 100 pages
of data on photoprocesses for ruthenium(II) complexes with polypyridyl ligands.
In many cases there is data at both liquid nitrogen (77 K) and room temperature
(298 K).  This is enough to obtain an ``average activation energy'' $E_{\mbox{ave}}$
using the equation,
\begin{equation}
  k(T) = A_{\mbox{ave}} e^{-E_{\mbox{ave}}/RT} \, .
  \label{eq:luminescence.2}
\end{equation}
We expect that this will underestimate the true $^3$TS activation energy 
($E_{\mbox{ave}} = \mbox{132 cm$^{-1}$}$ for [Ru(bpy)$_3]^{2+}$) but we do expect
it to preserve general trends.  Moreover it is easier to analyze one parameter than
have to deal with all six parameters that appear in the empirical formula of
Ref.~\cite{JBB+88}. 
In the end of our analysis of the experimental results in Ref.~\cite{JBB+88},
we have results for about a hundred complexes.

\begin{figure}
\begin{center}
\includegraphics[width=0.6\textwidth]{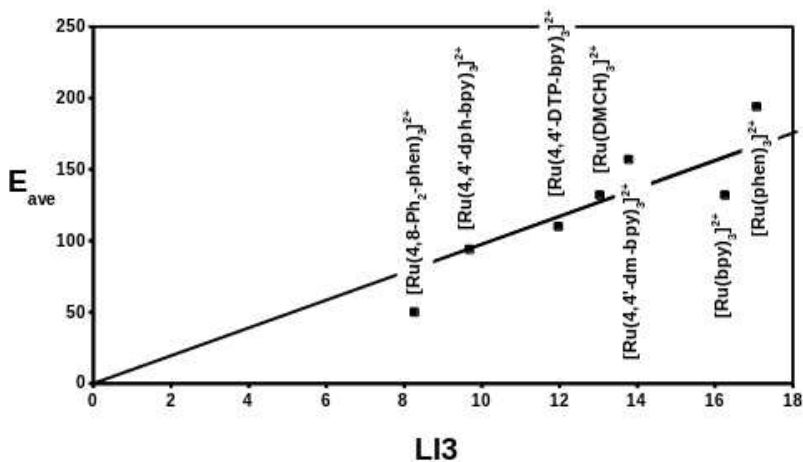}
\end{center}
\caption{
Performance of luminescence index number 3 for complexes of formula [RuX$_3$]$^{2+}$.
Both LI3 and $E_{\mbox{ave}}$ are in cm$^{-1}$.
Adapted from Ref.~\cite{MCA+17}.  See this same reference for a definition of the
ligand names.
\label{fig:allsameLI3}
}
\end{figure}
These results could be compared against calculated $^3$TS activation energies.  However
calculating $^3$TS activation energies for a hundred complexes is a formidable task.
It also may be difficult to explain the results back in terms of a LFT-like model.
We therefore opted for a less ambitious approach, namely to build a LI based upon
LFT-like orbital energies extracted from calculated PDOSs.  Different LIs were proposed
in Refs.~\cite{WJL+14,MCA+17}.  The one (LI3) that was found to work best \cite{MCA+17}
was not the most obvious choice but rather is based upon qualitative frontier-orbital-like 
reasoning,
\begin{equation}
  \mbox{LI3} = \frac{{\bar{E}}^2}{\Delta E} \, ,
  \label{eq:luminescence.3}
\end{equation}
where $\Delta E = \epsilon_{e_g^*} - \epsilon_{\pi^*}$ and 
$\bar{E} = (\epsilon_{e_g^*} + \epsilon_{\pi^*})/2$. 
Figure~\ref{fig:allsameLI3} shows that LI3 is an excellent predictor of $E_{\mbox{ave}}$
for polypyridine ruthenium(II) ligands composed of three identical bidentate ligands.
The picture is not so clean when we include other polypyridine ligands \cite{MCA+17},
although we are quite happy with the performance of such a simple
model, especially given the large number of approximations needed to extract $E_{\mbox{ave}}$.  
In particular, complexes with tridentate ligands appear to lie on a different line than 
complexes with bidentate ligands.


\section{Greener Alternatives}
\label{sec:green}

This section discusses greener alternatives to ruthenium in DSSCs and to iridium
in OLEDs.  Luminescence useful for other photodevices is also considered.  We first
consider replacing of rare expensive metals with more less expensive abundant metals
and then go on to considering how to eliminate metals entirely.

\subsection{Earth-Abundant Transition-Metal Complexes}
\label{sec:lightmetal}


\begin{table}
\begin{center}
\begin{tabular}{lccc}
\hline \hline
Compound 
         & $\lambda_{\mbox{abs}}$  
         & $\lambda_{\mbox{em}}$  & $\tau$  \\
\hline
ruby$^a$
         & 407 nm & 694 nm & 4.268 ms \\
$[$Cr(ddpd)$_2]^{3+,b}$
         & 435 nm & 778 nm & 898 ns \\
$[$Ru(bpy)$_3]^{2+,c}$
         & 452 nm & 615 nm & 950 ns \\
$[$Ir(ppy)$_3]^d$
         & 375 nm & 514 nm & 1.9 $\mu$s \\
\hline \hline
\end{tabular}\\
$^a$Al$_2$O$_3$(s) doped with Cr$^{3+}$ \cite{DB07},
$^b$In water \cite{OGF+15},
$^c$Refs.~\cite{NS11,AM16},
$^d$Refs.~\cite{KSW85,AM16}.
\caption{
\label{tab:LumiData}
Examples of luminescent compounds.
}
\end{center}
\end{table}

Luminescence is by no means restricted to complexes of rare expensive metals,
but it is only within the last decade or less that attention has turned to making
more efficient photodevices from less expensive Earth-abundant metals, including 
chromium, iron, copper, zinc, zirconium, molybdenum, and uranium \cite{LW18}.  
Although most
of the work has been done on copper(I) complexes, for reasons of space, we have chosen
to focus on the story of how a ``molecular ruby'' was made by ``rational ligand design''
\cite{OGF+15} through a close collaboration between 
researchers from coordination chemistry, physical inorganic
chemistry, and spectrocopy \cite{ONE+17}.

Ruby is a form of aluminum oxide (Al$_2$O$_3$) called corundum with additional 
chromium
impurities (Ref.~\cite{J06}, pp.~159-161, see also Ref.~\cite{J00}.)  
Cr$^{3+}$ ion replace Al$^{3+}$ ions 
in octahedral sites and are responsible for the well-known red color of ruby.  There
are three particularly important states which we will describe using LFT \cite{FH00}.  
In order of decreasing energy, these are,
\begin{equation}
  ^4T_2 : \underbrace{[\uparrow][\uparrow][\,\,]}_{t_{2g}}
          \underbrace{[\uparrow][\,\,]}_{e_g} \, \, , \, \, 
  ^2E : \underbrace{[\uparrow][\downarrow][\uparrow]}_{t_{2g}}
          \underbrace{[\,\,][\,\,]}_{e_g} \, \, , \, \, 
  ^4A_2 : \underbrace{[\uparrow][\uparrow][\uparrow]}_{t_{2g}}
          \underbrace{[\,\,][\,\,]}_{e_g} \label{eq:green.1}
  \, ,
\end{equation}
where the electron configuration from which each state is derived by symmetry projection 
is shown.  Ruby luminesces at 694 nm with a long luminescence lifetime of 1.8 $\mu$s
\cite{DB07} (Table~\ref{tab:LumiData}.)  This may be regarded as phosphorescence because
the intial spin-conserving photoexcitation $^4A_2 \rightarrow ^4T_2$ leads to a radiationless
intersystem crossing $^4T_2 \rightarrow ^2E$ with spin flip.  The final luminescence comes
from the $^2E \rightarrow ^4A_2$ transition which is nominally spin forbidden, leading to
a long luminescence lifetime. This luminescence was exploited in making the
first laser \cite{M60} (see also Ref.~\cite{T99}, pp.~104-106.)

Now suppose we want to make an octahedral Cr(III) coordination complex with properties
analogous to, but ideally even better than, ruby by varying the ligands around the central
metal atom.  How should we proceed?  It turns out that the main problem is the photoluminescence 
yield $\Phi$ [Eq.~(\ref{eq:Denis.1})], namely the number of photons that luminesce in comparison to the number of
photons that absorb.  $\Phi$ is low in Cr(III) complexes with low ligand field strength because
the energy ordering of the states is, $\mbox{\em Energy: } ^2E > ^4T_2 > ^4A_1$.  As the the $^4T_2$ state is
close in energy to the $^4A_1$ state, the ``energy gap law'' which states that radiationless
decay becomes more likely the closer an excited state is in energy to the ground state tells
us to expect that $\Phi$ will be very small.  For conventional strong-field ligands, 
$\mbox{\em Energy: } ^2E \approx ^4T_2 > ^4A_1$, which is also not good because the radiative $^2E$ and the
nonradiative $^4T_2$ states readily exchange energy, hence lowering $\Phi$.  Instead, we should
seek a very strong-field ligand so that, $\mbox{\em Energy: } ^4T_2 > ^2E > ^4A_1$, if we want to maximize $\Phi$.

The ligand that was found to satisfy these criteria gave a ``molecular ruby'' 
{\em bis}(N,N'-dimethyl-N,N'-dipyridine-2-ylpyridine-2,6-diamine)chromium(II) cation, 
[Cr(ddpd)$_2$]$^{3+}$ [Fig.~\ref{fig:LEDs}(b)]. 
(Ref.~\cite{FDR+18} provides a comprehensive review of the ddpd ligand and
its zinc, copper, nickel, cobalt, iron, chromium, and vanadium complexes.)
Molecular ruby turns out to be very stable, water soluble, and to have a luminescence lifetime of 899 $\mu$s
with a photoluminescence yield of 11\% \cite{OGF+15} (Table~\ref{tab:LumiData}.)
It may be used to produce singlet oxygen \cite{ONE+17}, create an optical molecular thermometer
\cite{OSB+17}, and as an alternative to ruby in measuring pressure in high pressure experiments
\cite{OHHR18}.
Further improvement of the photoproperties of molecular ruby may be obtained by 
deuteration, giving a luminescence lifetime of 2.3 ms and a luminescence yield of
30\% \cite{WOD+18}.

Let us now take a closer look at how computational chemistry helped in the design of molecular ruby.
We should not forget that the starting point was LFT, which may be regarded as one of the tools in
the computational chemists' toolbox. In fact, explicit LFT calculations were reported in
Ref.~\cite{OHHR18} as well as DFT and complete active space self-consistent field (CASSCF)
calculations.  LFT
tells us that we may increase the LFT splitting by chosing
a strong-field ligand with a large bite angle, guaranteeing a rigid octahedral geometry, and by 
making sure that it is a strong $\sigma$ donor.  Note that a rigid geometry will also
help to reduce vibrations
which are associated with radiationless relaxation.  
The ddpd ligand was designed computationally with the aim in mind of synthesizing another complex, related to, but different from, molecular ruby \cite{BFH10}.  Not every complex designed by computational chemists may be synthesized.  In fact, the initially
imagined complex could not be synthesized in this case \cite{BFH10}.  However that 
investigation also included a computational investigation of [Ru(ddpd)$_2$]$^{2+}$,
showing that the {\em mer} isomer should be more stable than the {\em fac} isomer and
that the bite angle of ddpd should be close to 90$^\circ$.  (In fact, the bite angle
reported in Ref.~\cite{BFH10} is 88$^\circ$.)
The bite angle in molecular ruby is in the range off roughly 
85$^\circ$ - 87$^\circ$ (Table 1 of Ref.~\cite{FDR+18}.)
Once the compound was synthesized, it was necessary to characterize the product to prove that it is
the correct compound.  In this case, X-ray crystallography could prove that the compound was the
correct one and photophysical methods could verify the spectra and photoluminescence yield, but 
computational chemistry is needed in order to validate the theoretical reasoning accounting for
the photophysical properties.  In Ref.~\cite{OGF+15}, the theoretical reasoning was validated by
using TD-DFT to assign the excited states, the calculation of excited state spin densities to 
validate that the chromium remains in the +III oxidation state, and the geometries of some of the
excited states were optimized in order to understand excited-state relaxation effects.
Ref.~\cite{ONE+17} includes a report on the results of a DFT study of the reaction of
$^1$O$_2$ with a model amine N(CH$_2$CH$_3$)(CH$_3$)$_2$.

\subsection{Organic Dye-Sensitized Solar Cells}
\label{sec:ODSSC}


As we have seen, one of the best dyes for use in DSSCs is
triscarboxy-ruthenium terpyridine [Ru(4,4',4"-(COOH)$_3$-terpy)(NCS)$_3$],
popularly known as the ``black dye.''  However ruthenium is expensive
(about 250 USD/ozt) and mildly toxic \cite{ZGC+17}. 
This has led to the desire to replace
the black dye (and other heavy-metal containing dyes) by metal-free
DSSC dyes.  We give a brief review of metal-free DSSC dyes 
in this section.  Some of these dyes are simply extracted from natural 
sources.  Others are 
``engineered.''  We follow up this 
literature review by an example from our own work of how computational 
chemistry can help in DSSC dye design.


Natural dyes containing carboxylic acid groups can easily react with the 
surface of TiO$_2$ nanoparticles.  Such dyes can often be extracted from 
plants, using green chemistry, without tedious synthesis, using relatively
simple separation and purification methods.   The resultant dyes 
may sometimes be used
as metal-free dyes in functioning DSSCs.  For example, in 1997, Gr\"atzel and 
co-workers were among the first to report the use of natural dyes in a DSSC 
by using blackberry juice to create a working DSSC \cite{CSGZ97}.  It is now 
even possible to find {\tt youtube} videos explaining how to make your own 
DSSC using blackberry juice \cite{DSSC}.   In recent years, the list of 
natural metal-free dyes for making DSSCs has considerably expanded.
(References~\cite{N12,LMM+14,KLL+17,RKTG17} review the state of the natural dyes
for use in DSSCs.)  Unfortunately, typical
PCEs are very low (less than 1\%).
%
Hence, molecular-level engineering to create better MF DSSC dyes seems quite
warranted.  In this context, computational chemistry modeling can hope to
provide a molecular-level understanding which can help in optimizing 
photoefficiency.  Indeed, there are many beautiful 
computational chemistry studies in the literature, including (but not limited to)
Refs.~\cite{DFG11,DG11,GD11,DMG12,LBK12,LLL+15,ZHS+18,CTH+19,SLL+19,SOL+19}.  
Let us just mention an explicit solvent molecular dynamics/TD-DFT study of 
the color of alizarin bound to TiO$_2$ as a particularly detailed 
study of a particular system \cite{DMG12}.  However, rather than study a 
specific system in detail, most studies seek to understand trends for 
series of candidate metal-free DSSC dyes.

A particularly interesting strategy used in the design of metal-free DSSC
dyes is to make push-pull molecules of the form D-$\pi$-A where D is an 
electron donor, $\pi$ is a conjugated spacer which serves as a molecular
wire for conducting electrons to A, the electron acceptor.
The push-pull structure reduces the risks of donor-acceptor electron 
recombination by promoting charge transfer and also accelerates 
the regeneration of the dye by the electrolyte.  The push-pull strategy
has been discussed for zinc porphorins in Ref.~\cite{LLW18} while
Ref.~\cite{SG19}
which studies the best spacers and the best length of spacers for $p$-type DSSCs.
Other articles concerning the push-pull strategy in designing metal-free
DSSC dyes include, but are not limited to, 
Refs.~\cite{LWL+08,XLP+08,LLS+09,ZBC+09,ZBL+09,CTW+11,JNHW11,NFE+12,MYG+14,XWBZ15,ZYS+15,%
BDG16,NYL16,LWSZ17,JZK18,HXM+19,LLXL19,SG19b,ZJK+19}.


Yet another strategy for making metal-free DSSC dyes is seen in 
theoretical attempts to create organic dyes
which will promote singlet fission (SF) \cite{PONT19}.  In SF, an energetic 
photon is absorbed creating a singlet state which can subsequently split into
two independent singlet or triplet excitations.  These excitations can
each interact with the TiO$_2$ substrate to produce two electrons.
PCE increases because for each photon absorbed two electrons
are produced.

\begin{figure}
\begin{center}
\includegraphics[width=0.5\textwidth]{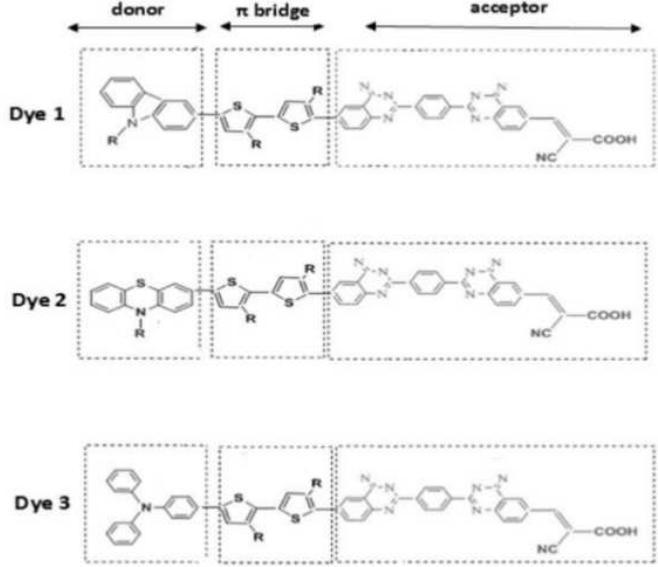}
\end{center}
\caption{
Chemical structures of the three push-pull dyes studied here, where R stands for 
the methyl group (CH$_3$).  The three dyes have the same $\pi$-spacer and acceptor unit, but differ by the donor unit:
methylcarbazole for dye {\bf 1}, 10-phenothiazine for dye {\bf 2}, and triarylamine for dye {\bf 3}.
\label{fig:ODSSCdyes}
}
\end{figure}
\begin{figure}
\begin{center}
\includegraphics[width=\textwidth]{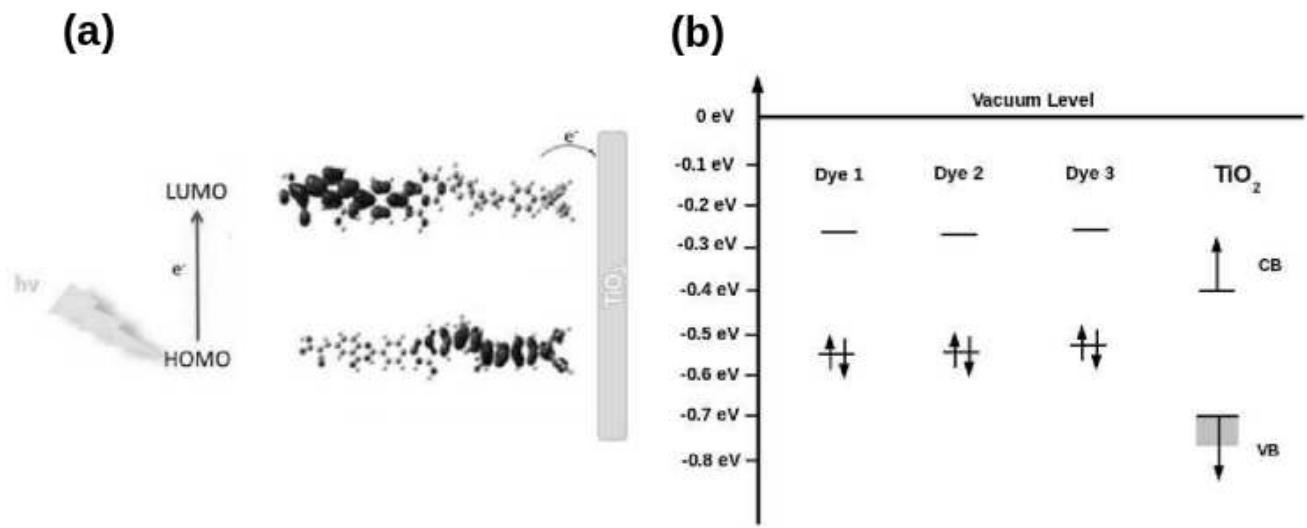}
\end{center}
\caption{
(a)
Calculated DFT HOMO and LUMO densities for dye {\bf 3} illustrating the HOMO $\rightarrow$ LUMO 
charge transfer that is expected to occur during the passage of an electron from the ground 
to the lowest excited singlet state.
(b)
Comparison of $E^{\mbox{dye}}$ and $E^{\mbox{dye}*}$ for the three dyes with the
TiO$_2$ valence band (VB) and conduction band (CB) level.
\label{fig:Tarek}
}
\end{figure}

We now we wish to provide a simple example of how computational chemistry
may be used in analyzing a series of candidate metal-free DSSC dyes \cite{M19}.
To this end, we consider the three ``push-pull'' molecules based upon
2,2'-(1,4-phenylene)bis(quinazolin-4(3H)-imine) shown in 
Fig.~\ref{fig:ODSSCdyes}.  Each of these molecules consists of an electron
acceptor A containing a carboxylic acid group that can bind to TiO$_2$
nanoparticles, a donor group D, and a $\pi$-bridge spacer group that also
serves as a ``molecular wire'' for transfering electrons from D to A.
Only D varies in this series of molecules while the $\pi$-bridge and A
is held constant.  We anticipate that a photon will create a charge separation
creating a hole on D and an electron on A.  This is consistent with the
picture shown in Fig.~\ref{fig:Tarek}(a).  We are interested in factors
that increase the PCE.  From Eq.~(\ref{eq:intro.2}), this includes $V_{oc}$
(also referred to as $\Delta G_{\mbox{inject}}$ in this field) which is
related to the difference between $\epsilon_{\mbox{HOMO}}^A = \mbox{-4.0 eV}$
for TiO$_2$ \cite{AWH+01} and $\epsilon_{\mbox{LUMO}}^D$ for the dyes and
may be thought of as the driving force for charge separation. 


\begin{table}
\begin{center}
\begin{tabular}{cccc}
\hline \hline
Dye & {\bf 1} & {\bf 2} & {\bf 3} \\
\hline
$E^{\mbox{dye}}$ & -5.472 eV & -5.454 eV & -5.261 eV \\
$E^{\mbox{dye$^*$}}$ & -2.692 eV & -2.704 eV & -2.611 eV \\
$\Delta G_{\mbox{inject}}$ & 1.308 eV & 1.296 eV & 1.389 eV \\
\hline \hline
\end{tabular}\\
\caption{
\label{tab:Tarek5}
Summary of some relevant energies for dyes {\bf 1}, {\bf 2}, and {\bf 3}.
}
\end{center}
\end{table}
The geometries were first optimized at the B3LYP/6-31G(d) level in vacuum.
The HOMO and LUMO energies were then obtained at the same level of calculation. All three
dyes have more negative HOMO energies than the I$^-$/I$^-_3$ redox
couple (-4.8 eV \cite{HG95}), indicating that a I$^-$/I$^-_3$ solution is
expected to be appropriate for quick regeneration of the ground state of the dye
after electron transfer to the substrate.  The ground state oxidation potentials of the
three dyes were obtained in the approximation, 
$E^{\mbox{dye}} = \epsilon_{\mbox{HOMO}}^{\mbox{dye}}$.  (Results are collected
in Table~\ref{tab:Tarek5}.) Spectra were calculated
at the TD-CAM-B3LYP/6-31G(d) level and the excitation energy
$\Delta E$ corresponding to the lowest energy absorption peak
and corresponding oscillator strength $f$ were obtained.  The oxidation potential
of the excited dye was calculated as 
$E^{\mbox{dye}*} = \epsilon_{\mbox{HOMO}}^{\mbox{dye}} + \Delta E$.
The resultant $E^{\mbox{dye}*}$ is from 0.27 eV to 0.38 eV higher in energy 
than the $\epsilon_{\mbox{LUMO}}^{\mbox{dye}}$ calculated at the B3LYP/6-31G(d) level.

Typically the charge-separation efficiency $\Phi_{\mbox{inject}}$ within the active
layer increases as $\Delta G_{\mbox{inject}}$ increases.  Figure~\ref{fig:Tarek}(b) 
part of the story: An electron initially in the HOMO of the dye is excited to the LUMO of the
dye. The calculated value of LHE given in Table~\ref{tab:Tarek5} shows that this is an 
efficient charge-transfer process with values of LHE close to unity.
As the LUMO of the dye is higher than the CB of the TiO$_2$, the electron jumps to the
CB.  The difference between the CB energy and the $E^{\mbox{dye}*}$ is related to the 
charge-separation momentum which will drive the electron into the TiO$_2$.  
As seen in Table~\ref{tab:Tarek5}, $\Delta G_{\mbox{inject}}$ is highest for dye {\bf 3}
and lowest for dye {\bf 2}, though the differences are not dramatic.
\subsection{Metal-Free Organic Light-Emitting Diodes}
\label{sec:MF-OLED}


In Sec.~\ref{sec:IrOLED}, we considered phosphorescent OLEDs that
exploit the heavy-metal effect by incorporating iridum(III) complexes in
their active layer.  There is also some work on using Earth-abundant
transition metals such as copper to make OLEDs \cite{YRC+11}.  However
we wish to ask the question of how metals might be removed from the
active layer and still have an efficient OLED?  We will follow the conventional
practice of calling this a metal-free (MF) OLED, though metals are always
still present in the electrodes.  
Here, as in the other areas treated in this chapter, a heavy synergy between 
experiment and computational chemistry is strongly felt in this area as 
computational modeling is important for clarifying luminescence mechanisms 
and for providing ideas for improved luminescent molecules.

Several strategies are possible for making MF-OLEDs, but only a few
will be mentioned here.  Thus MF-OLEDs may be based upon either 
phosphorescence or upon fluorescence.  Even in the MF case, OLEDs
may take advantage of the heavy-atom effect by incorporating non-metal
heavy atoms, such as bromine.  This is the case of a MF-OLED reported
by Lee, Han, and Lee \cite{LHL19}.  This study is especially notable
because the experimental work is complemented by TD-DFT calculations
characterizing the excited states and by molecular dynamics simulations
to study structure and rigidity.  It is also possible to make a 
phosphorescent MF-OLED without any heavy atom.  This is possible because
SOC is present, even though weaker than in the heavy-atom case, provided
non-radiative pathways can be blocked \cite{CSM+13}.  A drawback of this
latter strategy is that exploiting the long-lived phosphorescent state  
is not compatible with the high-driving intensities needed for
intense lighting \cite{CSM+13}.

A better strategy for making MF-OLEDs is to base them on fluorescence
rather than phosphorescence.  There are at least two ways to do this.
Molecules with a small enough $S_1$-$T_1$ gap can show thermally-activated 
delayed fluoresence (TADF).  This simply means that the meeting of
electrons and holes in the active layer produces both singlets and
triplets but that the Boltzman factor for thermal population of the
singlet state from the triplet state is large enough that the triplet
population drains away into the more rapidly luminescing singlet state. 
This is the strategy for the Earth-abundant copper complex-based
OLED mentioned above \cite{YRC+11} but has also been used in MF-OLEDs
\cite{UGS+12,KWC+16}.  A detailed theoretical mechanism for TADF in
conjugated polymers is reported in Ref.~\cite{BSPB01}.

Yet another strategy for making MF-OLEDs, and one that we choose to focus on here,
is to make use of stable organic free radicals.
Although the existance of stable organic free radicals does not seem to be
widely known in the chemistry community outside of specialists in this area,
stable organic free radicals have been studied since the 1900 landmark paper of
Gomberg on the triphenylmethyl radical \cite{G1900}.  Since then many other
stable organic free radicals have been discovered (Refs.~\cite{K83,H17} provide
useful reviews of different classes of stable organic free radicals).  
Recently stable organic radicals have been suggested as a way to make MF-OLEDs 
\cite{HKN14,HKN15,POZL15,OAZL16,AED+18,GPC+19}.

\begin{figure}
\begin{center}
\includegraphics[width=0.4\textwidth]{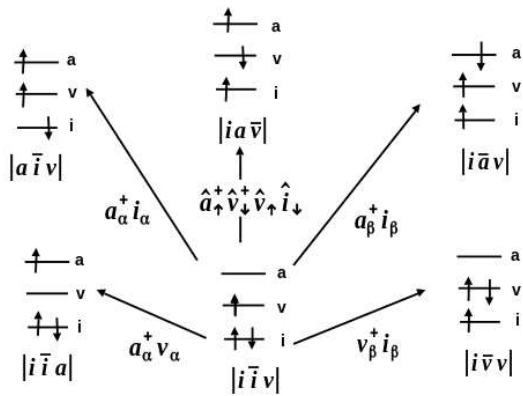}
\end{center}
\caption{
Three-hole three-electron model (TOTEM).  See text for additional explanation.
\label{fig:TOTEM3}
}
\end{figure}
In order to understand how this works, let us consider the TOTEM again, but
this time, it is the three-electron three-level model shown in 
Fig.~\ref{fig:TOTEM3}.  
This figure is to be understood as having been constructed using the 
spin-restricted open-shell Hartree-Fock (ROHF) method \cite{R60} [whose DFT
equivalent is the spin-restricted open-shell Kohn-Sham (ROKS) method \cite{RMH94}],
which basically may be understood in this case as using an orbital hamiltonian averaged
over $\alpha$ and $\beta$ spins.

We are now free to study the excitation energies of such a system.  The two lowest
excited states are expected to arise from the $i \rightarrow v$ and $v \rightarrow a$
excitations, with the later expected to be of lower energy than the former if the
general principle is respected that higher-lying orbitals have less orbital-energy
separation than do lower-lying orbitals.  This leads to two types of doublets, each
arising from one of the two configurations, $D_1 = \, ^2( v,a)$ and $D_2 = \, ^2(i,v)$.
A detailed analysis (not shown here lack of space) leads to the conclusion that
the $D_1$ excitation energy may be negative if the orbital energy difference is
less than the corresponding coulomb repulsion integral,
$\epsilon_a - \epsilon_v < [aa \vert vv]$.  Thus the HOMO of the GS may be
higher than the LUMO of the GS or, to use the language of solid-state physics, 
we may have a ``hole below the Fermi energy.''

In principle, free radical MF-OLEDs could avoid the triplet/singlet spin-statistics
problem because only doublet states are available at low excitation energies.
Fluoresence would then occur from these to the ground state.  Of course
these ideas must be confirmed by experiment as other scenarios might be 
possible. 
For example, injection of a hole
into the active layer before injection of an electron could conceivably lead to the
formation of singlet and triplet intermediates. 
Fortunately a detailed experimental and theoretical study shows that
this is not the case for at least one organic free radical MF-OLED \cite{AED+18}. 
We also note that
the predicted possibility of a hole below the Fermi level has also been 
found \cite{GPC+19}.  


%
%
%
%
%
%
%
%
%
%
%
%
%
%
%
%
%

\section{Conclusion}
\label{sec:conclude}


This chapter has focused 
on some of the greenest areas of the energy sciences, namely the use of abundant
solar energy and energy-efficient lighting and displays through light-emitting
diodes.  We have focused specifically on dye-sensitized solar cells and upon organic
light-emitting diodes.  Even this deliberately-chosen narrow scope has still left us
with a very large area to review.  Nevertheless we have made an effort to explain
the basics of how devices work, how their efficiency is measured in practice, 
the key underlying photoprocesses.  We have also taken pains to illustrate 
how computational chemistry is used hand-in-hand with experimental work to make
better DSSCs and OLEDs by first considering conventional DSSCs and OLEDs based upon
the heavy-metal effect, then by considering the replacement of rare metals with 
Earth-abundant metals, and finally by examining strategies for making metal-free
DSSC dyes and OLED active layers.  As might be expected from such an ambitious
plan, we are deeply aware of what has not been treated and of what has only been
treated shallowly.  Nevertheless we hope and trust that this modest chapter will be
helpful in showing how computational chemistry is actively helping to make the 
energy sciences more efficient, more sustainable, and more environmentally-friendly.

\section*{Acknowledgement}
\label{sec:thanks}

DM and MEC gratefully acknowledge helpful funding from the French Embassy in
Kenya and from the African School on Electronic Structure Methods and Applications
(ASESMA).  MT and KA gratefully acknowledge funding from the Government of 
Tunisia. Too many people have helped us over the years towards a better understanding
of the topic of this chapter to be able to name them all here, but particular thanks
go to the following for helpful discussions: 
Katja Heinze, Damien Jouvenot, Max~Lat\'evi Lawson~Daku, 
Fr\'ed\'erique Loiseau, Cleophas Muhavini Wawire, and Xiuwen Zhou.
The authors are grateful for computational resources provided by
the {\em Plateau du Centre d'Exp\'erimentation et de Calcul
Intensif en Chimie} ({\em PCECIC}) of the {\em Institut de
Chimie Mol\'eculaire de Grenoble} ({\em ICMG}) at the {\em
Universit\'e Grenoble Alpes} ({\em UGA}) where some of the calculations
reported here were carried out.

\section*{Author Contributions}
\label{sec:contribute}

MEC wrote sections~\ref{sec:intro} and \ref{sec:conclude} and 
subsections~\ref{sec:lightmetal} and \ref{sec:MF-OLED}.
DM wrote section~\ref{sec:luminescence} and subsections~\ref{sec:tradmod},
\ref{sec:PMD}, and \ref{sec:IrOLED}.
TM and KA wrote subsection~\ref{sec:ODSSC}.
MEC coordinated the writing and was responsible for editing.

\newpage
\bibliographystyle{myaip}
\bibliography{refs}
\end{document}